\documentclass[11pt,a4paper]{article}
\usepackage{color}
\usepackage{amsfonts}
\usepackage{amssymb}
\usepackage{graphicx}
\usepackage{amsmath}
\usepackage[dvipdfm]{hyperref}
\usepackage{enumerate}
\RequirePackage{mathrsfs} \RequirePackage[sc]{mathpazo}
\RequirePackage{wasysym} \RequirePackage{setspace}
\usepackage{color}
\makeatletter \@addtoreset{equation}{section}

\newcommand{\be}{\begin{equation}}
\newcommand{\ee}{\end{equation}}
\newcommand{\bea}{\begin{eqnarray}}
\newcommand{\eea}{\end{eqnarray}}
\begin{document}

\title{ On  Thermodynamics of  $2d$ Black Holes in  Brane Inflationary Potentials }
\author{  A. Belhaj$^{1,2}$, M. Chabab$^{2}$, H. El Moumni$^{2}$,  M. B. Sedra$
^{3}$,  A.  Segui$^{4}$ \\
\\
{\small $^{1}$ D\'epartement de Physique, Facult\'e
Polydisciplinaire, Universit\'e Sultan Moulay Slimane}\\{
\small B\'eni Mellal, Morocco} \\
{\small \hspace{-1cm} $^{2}$ Laboratoire de Physiques des Hautes
\'{e}nergies et Astrophysique, FSSM, Universit\'{e} Cadi Ayyad}\\{
\small Marrakech, Morocco } \\
{\small $^{3}$  LHESIR,   D\'{e}partement de Physique, Facult\'{e}
des Sciences, Universit\'{e} Ibn Tofail }\\{ \small K\'{e}nitra,
Morocco}\\ {\small $^{4}$ Departamento de F\'isica Te\'orica,
Universidad de Zaragoza, E-50009-Zaragoza, Spain} }
\date{}
\maketitle

\begin{abstract}
 Inspired from  the inflation brane world cosmology, we study   the  thermodynamics of a
black hole solution  in two dimensional  dilaton gravity with an
arctangent potential background.  We first derive the  two
dimensional black hole geometry,  then we examine its asymptotic
behaviors.  More precisely,   we  find  that such behaviors exhibit
properties appearing  in some   known cases including the  Anti de
Sitter and  the Schwarzchild black holes. Using the complex path
method, we compute the Hawking radiation. The entropy  function can
be related to
the value of the potential at the horizon.
\end{abstract}

%\newpage
%\tableofcontents \thispagestyle{empty} \newpage \setcounter{page}{1}
\newpage

\newpage
\section{Introduction}

In the last years,   the black hole  physics   has received an
increasing attention in the context of supergravity theories
embedded in superstrings and M-theory compactified on Calabi-Yau
manifolds[1-8].   A  particular emphasis  has placed  on   the study
of extremal black solutions  in  various dimensions. This concerns
the attractors of  extremal black holes and branes. In this way, the
corresponding effective potentials and entropy functions have been
computed in terms of  the U-duality black brane charge invariants
\cite{www}.

 More recently, a  special  effort  has been  devoted to study  the  black
hole  solutions  in two-dimensional models of  the dilaton gravity
\cite{1,001,2,002}. This black hole solution   has been subject to
some interest not only because of its connection with the conformal
field theories but also from its connection with the  cosmological
models embedded in heterotic M-theory theory compactification in the
presence of  the dilaton coupling functions \cite{3,4}. In such
models, the Bekenstein-Hawking entropy has been  investigated and
explicitly calculated. The black hole thermodynamics  in two
dimensional dilatonic models have been also studied by several
investigation groups\cite{006,0006,5,6}. It has been shown, in many
places, that the black hole thermodynamics can open gates to
understand the quantum features of gravity theory.  It is possible
to explore such thermodynamic properties to study certain field
theory models using recent string theory technics including AdS/CFT
conjecture.

On the other hand, the F(R)-gravity theories, in the presence of a
scalar field,  have been extensively studied in the connection with
cosmological models  embedded in  superstring  models   and
M-theory\cite{007,600,601}. In particular,  a  new   F(R)-gravity
theory with a Lagrangian density involving an arctangent  function
has been studied in \cite{602}. The corresponding   static solutions
and the potential of the scalar field  have been obtained, showing a
good agrement with the PLANCK data.

The aim of this work is to contribute to these activities by
presenting  a black hole solultion in   two dimensional dilaton
gravity in the presence of  an  inflationary potential. More
precisely, we consider an arctangent potential background, explored
in the  brane world cosmology,  to  build  a   two  dimensional
black hole solution. Up to some limit conditions, it has been shown
that this black hole geometry  recovers some known results,
including the Schwarzchild and Anti de Sitter  black holes. Using
the  complex path analysis, the Hawking radiation is computed and
commented  in terms of   known black hole solutions. Then we
calculate the entropy function. This function can be related to the
value of the potential at the horizon.
\section{Two dimensional black hole in   brane inflationary  potential}
In this section, we  study   two dimensional black hole solutions in
the presence of an effective   potential used in brane world
%inflation models. To get that, we first consider an action
%describing  a non-dynamical scalar field coupled to the Ricci
%scalar. This  action reads as
%\begin{equation}\label{1}
%\mathcal{S}=\int d^{2}x\sqrt{-g}(V(\phi)+D(\phi)R).
%\end{equation}

 To get   two singular two-dimensional dilatonic
black hole solutions, one should first  start  with the most
 general  lagrangian involved  in  the  dilatonic gravity in \cite{9212}
\begin{equation}\label{Lag}
\mathcal{L}=\sqrt{-g}(D(\phi) R+ G(\phi)(\nabla\phi)^2+H(\phi)).
\end{equation}
Then,  one  performs the following conformal transformation
\begin{equation}
g_{\alpha\beta}\rightarrow e^{2\sigma(\phi)g_{\alpha\beta}}
\end{equation}
with
\begin{equation}
\frac{d\sigma}{d\phi}\frac{dD}{d\phi}=-\frac{1}{4}G(\phi).
\end{equation}
In fact, the kinetic term $G(\phi)$   can  be  absorbed in the
redefinition of  the  $D(\phi)$ function. Indeed, redefining  the
potential as $V(\phi)=e^{2\sigma(\phi)}H(\phi)$, and transforming
the kinetic term, one gets  the action  appearing in \cite{9212}
\begin{equation}
\mathcal{S}=\int d^{2}x\sqrt{-g}(V(\phi)+D(\phi)R).
\end{equation}
It is worth noting that  the field equation for  the scalar field
$\phi$ obtained from this action   does not contain derivative
terms. In fact, this should be equivalent to  the action involving
the  higher derivative    with  only the gravitational field namely
\cite{9509}
\begin{equation}
\mathcal{S}_{HD}=\int d^2x \sqrt{-g}\mathcal{F}(R)
\end{equation}
 It turns out  that the lagrangian (\ref{Lag}) can be extended by introducing   $N$
conformally coupled scalar fields  $f_i$  \cite{anomal}
%[Phys.Rev.(53)8 1996]
\begin{equation}\label{matter}
\mathcal{L}=\sqrt{-g}(D(\phi) R+ G(\phi)(\nabla\phi)^2+H(\phi))-\frac{1}{2}\sum_{i=1}^N (\nabla f_i)^2.
\end{equation}
However, the corresponding  equations of motion are not easy to deal
with. For this  reason, we consider here  the simplest case without
conformal coupling matter fields.  The general  study might be
examined in future works using  sophisticated numerical analysis.

In connection with  the higher dimensional theories,  this action
could be obtained
 from    M-theory on nine dimensional compact
 manifolds. Indeed, within  heterotic  M-theory framework,  the
 internal space  can be identified with  the Calabi-You fourfolds
 fibred by the orbifold $S^1/Z_2$  obtained by implementing the $Z_2$ action on the circle
 coordinate. In this way, the corresponding models can be derived
 from a three dimensional  heterotic M-theory  with orbifold
 data \cite{7}.  In the  heterotic M-theory,  the models associated  with  this action
 can be
controlled by  the $V(\phi)$ and $D(\phi)$ functions  depending on
the  scalar  field known by the dilaton in string theory and related
models.

 Assuming that this action can be obtained from  the superstring compactification down to two dimensions
  and varying it  with respect to the dilaton  scalar field, one
gets
\begin{equation}\label{2}
-\sqrt{-g}\frac{\partial V}{\partial\phi} (\phi)=\frac{\partial
D}{\partial \phi} (\phi) \sqrt{-g}R.
\end{equation}
where $R_{\alpha\beta}=g_{\alpha\beta}\frac{R}{2}$. However, the
variation with respect to the metric tensor yields
\begin{equation}\label{5}
V(\phi)g_{\alpha\beta}=2(\nabla^2
g_{\alpha\beta}-\nabla_\alpha\nabla_\beta)D(\phi).
\end{equation}
In what follows, we  discuss a two-dimensional dilatonic  black hole
in a  special background   based on a  potential  explored  in the
inflationary brane physics. Precisely, we consider a spheric
symmetric static solution, in the Schwarzschild like gauge,  given
by
\begin{equation}\label{6}
ds^{2}=-f(r)dt^{2}+\frac{1}{f(r)}dr^{2},
\end{equation}
where $f(r)$ is a scalar function specified later one in terms of
the inflationary  cosmological date including the  potential. This
metric can be obtained generally from a $d$ dimensional  static
spherically symmetric solution
\begin{equation}\label{6}
ds^{2}=-f(r)dt^{2}+\frac{1}{f(r)}dr^{2}+r^2d\Omega^2_{d-2}
\end{equation}
using  the compactification on  the  $(d-2)$-dimensional sphere.

The  metric  function $f(r)$ is determined by fixing the $V(\phi)$
and $D(\phi)$ functions. Indeed, we take  a particular value of the
dilaton scalar field given $D(\phi)=\frac{1}{\phi}$. The above
equations can be reduced to
\begin{equation}\label{7}
\frac{\partial V}{\partial \phi}(\phi)=\frac{1}{\phi^2}R
\end{equation}
together with
\begin{equation}\label{8}
g_{\alpha\beta}V(\phi)=2(g_{\alpha\beta}\nabla^2
-\nabla_\alpha\nabla_\beta)\frac{1}{\phi}.
\end{equation}
Technical calculations  give  the
 following equations
\begin{eqnarray}\label{10}
\phi^3 V(\phi)+2 f\phi\phi^{''}-4 f \phi^{'2}+f^{'} \phi\phi^{'}
=0 \nonumber\\
\phi^2 V(\phi)+f^{' }\phi^{'}    =0\\
\frac{\partial V}{\partial\phi} (\phi)+\frac{1}{\phi^2}(f^{''} )
=0.\nonumber
\end{eqnarray}
Having fixed the $D(\phi)$ function, the model will depend now only
on the potential form. It turns out that there are many physical
potentials incorporated  in string theory and related models
including M and F-theories moving on orbifold fixed spaces \cite{7}.
Such potentials have been used in the moduli stabilization in
stringy cosmological models. In fact, the scalar potential shape
turns out to be essential in  the elaboration of stringy
inflationary models. The well studied  models  are   the chaotic
inflation potential reported in \cite{82,83} and the minimal
supersymmetric standard model inflation discussed in \cite{84,85}.
Besides such examples, there are several other models which have
been  also investigated in the literature.

To keep contact  with   the brane inflationary  models, we will
 consider a   cosmological potential satisfying  the limiting curvature
condition proposed in \cite{9}.  This  inflationary potential takes
the form
\begin{equation}\label{16}
V(\phi)=\lambda   \arctan(\phi ^2),
\end{equation}
where $\lambda$ is a constant.  This constant represents the large
field approximation.   It is worth noting that this potential has
many nice cosmological properties\cite{10}. In particular, for very
small values of the dilaton scalar  field,  it behaves as the
chaotic one. More precisely, when $\phi^2$ is very close to zero, at
second order,  a chaotic inflation scenario  can be recovered
\cite{82,83}. Moreover,   the Taylor expansion at order 10 can
reproduce  the MSSM inflation  potential studied in\cite{84,85}.

 After specifying the  above functions, we  are  now in position to derive  the
associated $f(r)$ metric  function  producing a two dimensional
black hole solution.   This can be done according  to the analysis
presented in \cite{13}. To
   present such  solutions, let us first take some conditions on the
dilaton  field given by $\phi=\frac {1}{\alpha r}$, where $\alpha $
is a constant. After calculation, we get
\begin{eqnarray}\label{fr}\nonumber
f(r)&=& C-\frac{C r}{r_0}+\frac{r\lambda}{\alpha}(\arctan[\alpha^2r_0^2]-\arctan[\alpha^2r^2])\\ \nonumber
&+&\frac{\lambda}{\alpha\sqrt{2}}(\arctan[1+\alpha \sqrt{2}r]-\arctan[1-\alpha \sqrt{2}r])+\frac{r\lambda}{\sqrt{2}r_0\alpha}(\arctan[1-\alpha \sqrt{2}r_0]-\arctan[1+\alpha \sqrt{2}r_0])\\
&+& \frac{\lambda}{2\sqrt{2}\alpha^2}Log\left[\frac{1-\sqrt{2}\alpha r+\alpha^2r^2}{1+\sqrt{2}\alpha r+\alpha^2r^2}\right]+\frac{r\lambda}{2\sqrt{2}\alpha^2}Log\left[\frac{1+\sqrt{2}\alpha r_0+\alpha^2r_0^2}{1-\sqrt{2}\alpha r_0+\alpha^2r_0^2}\right]
\end{eqnarray}
where $r_0$ is the location of  the black  hole horizon, and $C$ is
a constant of the  integration, fixed later one. The computed
function $f(r)$ is presented in figure 1.

 \begin{center}
\begin{figure}[!h]
\hspace{3 cm} {\includegraphics[scale=1]{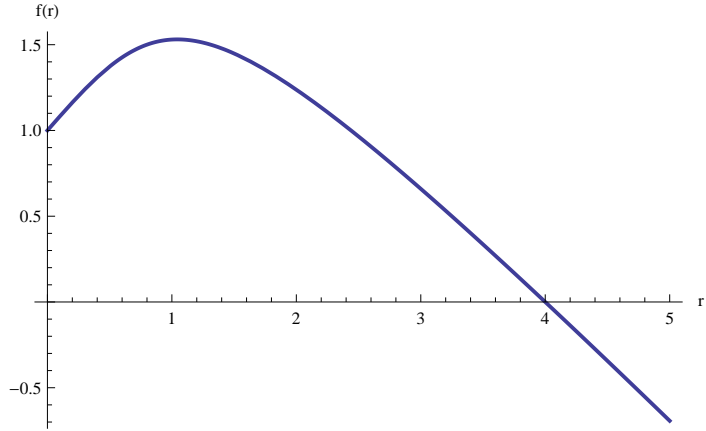}}
\vspace*{-.2cm} \caption{A plot of $f(r)$ for the values $\alpha=\lambda=C=1$, the function $f(r)$ vanishes at the horizon $r=r_0$.} \label{fig1:fig2}
\end{figure}
\end{center}

Before going ahead,  let us comment the obtained  solution. In fact,  we  will see that  the
 above potential  can recover  some  known black hole solutions in two
 dimensions  including Schwarzchild  and Anti
de Sitter ones. Indeed,   asymptotic regions in the moduli space for
either large or vanishing
 values of $\phi$    produce a   black hole
solution controlled by the following  solution
\begin{equation}
f(r)\underset{\alpha\to0,\infty}{\longrightarrow}C\left(1-\frac{r}{r_0}\right).
\end{equation}
 Such regions are defined  when
$\alpha$ goes to zero or infinity that mean for large  or vanishing values of $\phi$. To see that, it  is useful to
consider the following  radial variable transformation
\begin{equation}\label{trans}
 r\mapsto r^\xi
\end{equation}
where  $\xi$ is a natural number fixed  later one.  Putting    $\xi=2$,   we can  recover the
Anti de Sitter  metric solution. In this case, the  $f(r)$  function
tends to
\begin{equation}
f(r)=C\left(1-\frac{ r^2}{r_0^2}\right).
\end{equation}
Fixing   again the constant of the  integration to one ($C=1$)  and
identifying the event horizon  with  the cosmological constant
$r_0^2=-\ell^2$, the above equation  becomes
\begin{equation}
f(r)=1+\frac{r^2}{\ell^2}.
\end{equation}
This  particular  value of $\xi$  reproduces  the result  presented
in
\cite{1}. \\
Taking  $\xi=-1$,   we can  get  the  metric  describing the
Schwarzchild black hole. In this case, the function  $f(r)$ becomes
\begin{equation}
f(r)=C\left(1-\frac{r_0}{r}\right).
\end{equation}
As in the previous  case,    setting  $C$ to $1$ and identifying
$r_0$ with the mass of black hole $r_0=2m$, one gets
\begin{equation}\label{xxsch}
f(r)=1-\frac{2m}{r}.
\end{equation}
The second comment  that we should add concerns the  Ricci scalar
$R$. Using (2.6),   the  above  potential  produces the  following
Ricci scalar $R$
\begin{equation}\label{19curv}
R=\frac{2 \alpha  \lambda  r}{\alpha ^4 r^4+1}.
\end{equation}
 This quantity  is finite for all positive values of the radial coordinate
 $r$. Indeed, the corresponding function is plotted below in figure 2.
 \begin{center}
\begin{figure}[!h]
\hspace{3 cm} {\includegraphics[scale=1]{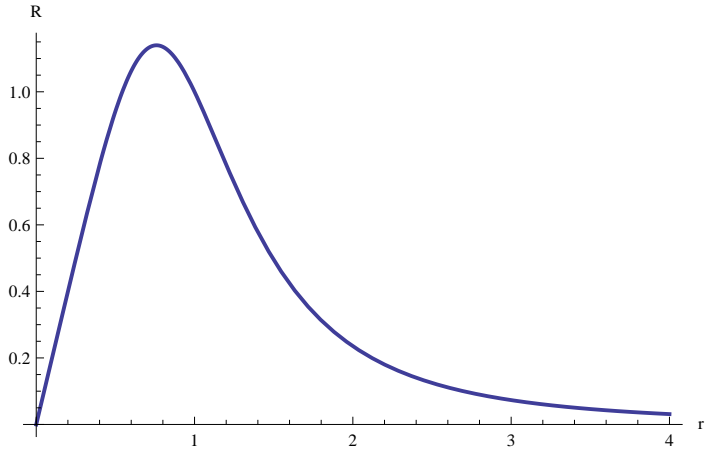}}
\vspace*{-.2cm} \caption{The curvature $R$ remains finite and positive for all $r$ from $r=0$ to $r=\infty$.} \label{fig1:fig2}
\end{figure}
\end{center}
It follows from this figure that the  curvature remains positive.
Moreover, the Ricci scalar function $R$  has a maximum at
$r=\frac{1}{\sqrt[4]{3} \alpha }$. The corresponding value    is
given by
\begin{equation}
\label{Rmax} R_{max}=\frac{1}{2} 3^{3/4} \lambda.
\end{equation}
The last comment  concerns  the mass of the obtained black hole.
Based on the result of \cite{10},  the $\lambda$ parameter seems to
behave as a   mass term  in the attractor mechanism, although it is
fixed at infinity and not at the horizon. This observation is
inspired from the analysis of  the four dimensional  brane inflation
model associated with the  potential given in (10). In this way, the
scalar large value  has been shown to be identified with the mass of
the  Schwarzchild black hole.   It is recalled that  the
Friedmann-Robertson-Walker equation in two dimensional
inflation-brane ($\cite{FRWxxx}$) background can be written as
\begin{equation}
\label{H}
H^2=\left(\frac{\dot{a}}{a}\right)^2=\left(\frac{\kappa}{2}\rho\right)^2-\frac{R+2m}{a^2}
\end{equation}
where $a(t)$ is the scale factor,  and where  $H(t)$ is  the Hubble
parameter.  In this equation,
 $\kappa$ is related to the D1-brane tension and the Planck mass.
 The parameter $R$ is the curvature of space given by $(\ref{19curv})$, and $m$ denotes   the mass of the black hole.
The dynamics can be described by a perfect fluid with a time
dependent energy density $\rho=\frac{1}{2}\dot{\phi}^2+\lambda
arctan(\phi^2)$ and the pressure
 $p=\frac{1}{2}\dot{\phi}^2-\lambda arctan(\phi^2)$ is the dominate energy distribution to  the  inflation.
 The dilaton  field satisfies the Klein-Gordon equation
\begin{equation}
\ddot{\phi}-3H\dot{\phi}+\frac{2 \lambda  \phi }{\phi ^4+1}=0.
\end{equation}
In the large  field approximation,  the effective  potential behaves
like  $V(\phi)\sim \lambda$. In this limit,  (\ref{H}) can be
reduced to
\begin{equation}
H^2=\left(\frac{\dot{a}}{a}\right)^2=\left(\frac{\kappa}{2}\rho^2\right)^2-\frac{2m}{a^2}.
\end{equation}
and the  Klein-Gordon equation becomes
\begin{equation}
\ddot{\phi}-3H\dot{\phi}=0.
\end{equation}
Solving the above  equation,  the expression of the dilatonic field
is given by
\begin{equation}
\phi(t)=\phi_\infty+\phi_0 \; exp(-3Ht).
\end{equation}
Asymptotic behaviors   require that  $\phi_\infty$  should  take
very large values. In this limit, the dilaton field is almost
constant and becomes very large when  $t$ goes to infinity.  In what
follows, one considers the  scale factor
\begin{equation}\label{scalfact}
a(t)=a_0 exp(H t).
\end{equation}
Taking a
particular geometry given by
\begin{equation}
H=\lambda
\end{equation}
and using (\ref{scalfact}), we get the following equation
\begin{equation}
m(t)\sim \lambda^2a_0^2exp(\lambda t).
\end{equation}
In the case where $\lambda$ takes small values,  that is in the limit $\lambda t<<0$, the mass equation becomes
\begin{equation}
m\sim \lambda
\end{equation}
where $a_0^2$ has been fixed to  $\frac{1}{\lambda}$. This equation relates the mass of the black hole and the asymptotic value of the potential.
\section{Thermodynamic properties }
 Having constructed  the
black hole solution,  a particular emphasis will  put on  discussing
its thermodynamic properties including  the Hawking radiation and
the  entropy function. First, we compute the Hawking radiation using
a  method based on the  complex path analysis introduced by Landau
\cite{land}. In order to derive that, it is essential to recall such
a  method. Assuming that  $f(r)$ vanishes at $r_0$ and $f'(r)$ is
nonzero at $r_0$ and expanding  it  around the point $r_0$, one
obtains
\begin{equation}\label{12}
   f(r)=f'(r_0)(r-r_0)+ \mathcal{O} \left[(r-r_0)^2\right]
    \equiv \mathcal{R} (r_0)(r-r_0).
\end{equation}
To show how to get the expression of the temperature,  it is useful
to consider an auxiliary  field $\Phi$ interacting with the above
system and satisfying the Klein-Gordon equation
\begin{equation}\label{13}
    \left(\square-\frac{m_0^2}{\hbar^2}\right) \Phi=0.
\end{equation}
Expanding this equation,  one finds  the following expression
\begin{equation}\label{14}
    -\frac{1}{f(r)}\frac{\partial^2\Phi}{\partial t^2}+\frac{\partial}{\partial
    r}\left(f(r)\frac{\partial\Phi}{\partial
    r}\right)=\frac{m_0^2}{\hbar^2}\Phi
\end{equation}
To get a  semiclassical wave function, one should
 consider   the  following standard ansatz
\begin{equation}\label{15}
    \Phi(r,t)=e^{\frac{i}{\hbar}\mathcal{S}(r,t)}.
\end{equation}
This ansatz  leads to
\begin{equation}\label{16}
    -\frac{1}{f(r)}\left(\frac{\partial\mathcal{S}}{\partial
    t}\right)^2 +f(r)\left(\frac{\partial\mathcal{S}}{\partial
    r}\right)^2+ m_0^2 - \frac{i}{\hbar}\left[\frac{1}{f(r)}\frac{\partial^2\mathcal{S}}
    {\partial t^2}-f(r) \frac{\partial^2\mathcal{S}}{\partial r^2}-\frac{df(r)}{dr}\frac{\partial\mathcal{S}}{\partial
    r}\right]=0.
\end{equation}
Expanding  $\mathcal{S}$ in a power series of $\hbar/i$ and
neglecting the higher orders,  one obtains
\begin{equation}\label{18}
\mathcal{S}_0=-\mathcal{E} t \pm \int
\frac{dr}{f(r)}\sqrt{\mathcal{E}^2-m_0^2f(r)}
\end{equation}
where $\mathcal{E}$ is a constant witch  can be  identified with the
energy.  The  desired formula can be derived by    taking   a simple
case corresponding to $m_0=0$.  Indeed, we  consider     the usual
saddle point method to calculate  the semiclassical propagator
$\mathcal{K}(z'',z')$ for a particle propagating from a spacetime
point $z''(t_1,x_1)$ to $z'(t_2,x_2)$. The propagation  takes the
form
\begin{equation}\label{19}
\mathcal{K}(z'',z')=\mathcal{N}\exp\left[\frac{i}{\hbar}\mathcal{S}_0(z'',z')\right]
\end{equation}
where  $\mathcal{N}$ is a normalization constant and $\mathcal{S}_0$
reads as
\begin{equation}\label{20}
    \mathcal{S}_0(z'',z')=-\mathcal{E}(t_2-t_1) \pm \int_{x1}^{x2}
    \frac{dr}{f(r)}.
\end{equation}
To compute  the amplitudes and the  probabilities of the  emission
and the  absorption  at $r_0$, one may consider an outgoing particle
at $x = x_1 < r_0$.  In this way, the  modulus squared of the
amplitude for such a  particle crossing  the horizon can  give the
probability of the  emissions.  Roughly,  the  complex analysis on
the plane reveals  the following expression
\begin{equation}\label{21}
    \mathcal{S}_0[emission]=\frac{i\pi \mathcal{E}}{\mathcal{R}(r_0)}+ \text{real
    part}, \qquad
   \mathcal{S}_0[absorption]=-\frac{i\pi \mathcal{E}}{\mathcal{R}(r_0) }+ \text{real
    part}
\end{equation}
Following \cite{1}, it has been shown that  the squaring  of the
modulus leads to
\begin{equation}\label{24}
P[emission]= \exp\left[\frac{-4\pi \mathcal{E}}{\mathcal{R}}\right] P[absorption]
\end{equation}
A close inspection  and  a comparison with    Hawking and  Hartle
formulation $
    P[emission]=e^{-\beta\mathcal{E}}P[absorption]$,  gives the
    following expression
\begin{equation}\label{26}
 T_H=\beta^{-1}=\frac{ |\mathcal{R}|}{4\pi}.
\end{equation}
Using equations($\ref{12}$) and  ($\ref{fr}$),  the temperature can
be written as
\begin{equation}\label{x1}
T_H=\frac{1}{4r_0 \alpha^2}\left|-4 C \alpha^2+\sqrt{2} \lambda \left( 2 arctan[1-\sqrt{2} r_0\alpha]-2 arctan[1+\sqrt{2} r_0\alpha]+\log\left[\frac{1+r_0\alpha(\sqrt{2}+r_0\alpha)}{1+r_0\alpha(-\sqrt{2}+r_0\alpha)}\right]\right)\right|
\end{equation}

 \begin{center}
\begin{figure}[!h]
\hspace{3 cm} \includegraphics[scale=1]{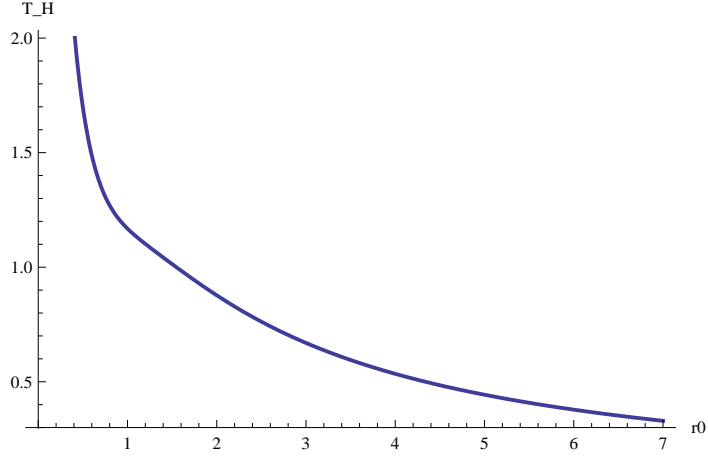}
\vspace*{-.2cm} \caption{A plot of Hawking temperature $T_H$ for the values $\alpha=\lambda=C=1$} \label{fig1:fig2}
\end{figure}
\end{center}

The temperature variable  allows one  to  define new   regions in
the spacetime and the moduli space. Applying an appropriate
approximation, we can recover some usual results. Indeed,  in the
region corresponding to $\alpha$ goes to infinity  or zero, the
above equation  reduces to the following form
\begin{equation}\label{x1}
T_H=\frac{1}{4\pi}\left|\frac{\text{C}}{r_0}\right|.
\end{equation}
 It follows from this equation that the constant $C$ controls
the black hole evaporation. This limiting behavior can recover the
expression   black hole. Identifying $r_0$ with the mass of the
Schwarzchild black hole $r_0=2m$, as we made before, we obtain
\begin{equation}\label{x1}
T_H=\frac{1}{8 \pi m}.
\end{equation}
It is observed   that for large  or small values of the dilaton
scalar field $\phi$,  the thermodynamical behavior remains  the
same. This can be interpreted as a possible duality between two
different regions associated  with the  small and large values of
the dilaton scalar field $\alpha$. It looks like the strong/weak
coupling duality observed in superstring theory.  This observation
merits an investigation. Moreover,  a two dimensional version of
Stefan's law  produces  the total power radiated by the black hole.
This can be  given  by the following relation
\begin{equation}
\mathcal{P}\sim - \frac{d m}{d t}\sim T_H^2
\end{equation}

It is worth noting that the entropy is one of the important
thermodynamic quantity of the  black hole physics. Indeed, we first
calculate   the surface gravity $\kappa$ evaluated at the event
horizon\cite{c11}. The result is
\begin{equation}
\kappa=-\frac{\lambda}{2}\arctan(\alpha^2 r_0^2).
\end{equation}
The variation of the energy  takes the following form
\begin{equation}
\delta \epsilon=-\frac{\lambda}{2 G}\arctan(\frac{1}{\alpha^2 r_0^2})\delta\left(\frac{1}{\alpha^2 r_0}\right)
\end{equation}
The first law of  the  thermodynamics $\delta\epsilon=T\delta S$
produces the following entropy function
\begin{equation}
S=\frac{2\pi\alpha^2 r_0^2}{G}.
\end{equation}
 The $\alpha^2 r_0^2$  quantity   denotes the value of the potential  at the black
hole horizon when $\alpha$ goes  to zero.

\section{Conclusion}
In this work, we have studied  a black hole solution  in two
dimensional dilaton gravity. To make contact with the  string
inflation brane world, we have dealt with a model based on an
arctangent potential background. Starting from this particular
potential form, we have built   first  a  two dimensional black hole
solution. Under some limit conditions, this    solution recovers
some known examples obtained  in  the literature.  Using the complex
path analysis, we have computed the  Hawking radiation. In
particular, we have considered some asymptotic   behaviors
reproducing the Schwarzchild  black hole case. The entropy function
has been also calculated and can be related to the value of the
potential at the horizon.

This work comes up with many questions related to possible  extended
potentials. A curios question  concerns the connection with   string
theory compactified   on  the Calabi-Yau manifolds producing extra
scalar fields associated with the  stringy moduli space.  In fact,
the potential used here could  be extended to
\begin{equation}\label{17}
V(\phi_i)=\sum_i\lambda_i   \arctan(\phi_i^2).
\end{equation}
 In this way, these scalar  fields can  be identified with the R-R B-field on 2-cycles
of the Calabi-Yau manifolds.  Open  string field axions,  appearing
in the supersymmetric consistent  D-brane models,  can be also
implemented in the discussion. Usually, the letters are related with
the Peccei-Quinn symmetry in four dimensions.  In connection with
higher  dimensions, it should be also interesting to go up by
considering four dimensional  models  and make contact with the
black hole  attractor mechanism and cosmological activities
\cite{9712}.

{\bf Acknowledgments}: AB would like to thank  P. Diaz and M. Naciri
for discussion on the related topics.  AS is supported by the
Spanish MINECO (grants FPA2009-09638 and FPA2012-35453) and
DGIID-DGA (grant 2011-E24/2).


\begin{thebibliography}{99}

\bibitem{V1} H. Ooguri, A. Strominger, C. Vafa, {\em Black Hole Attractors and the
Topological String}, Phys. Rev. {\bf D70} (2004) 106007, {\tt
hep-th/0405146}.
\bibitem{V2} C. Vafa, {\em Two dimensional Yang-Mills, black holes and topological
strings}, {\tt hep-th/0406058}.
\bibitem{V3} M. Aganagic, A. Neitzke, C. Vafa, {\em  BPS
Microstates and the Open Topological String Wave Function}, {\tt
hep-th/0504054}.
\bibitem{B1} A. Belhaj, P. Diaz, A. Segui, {\em Magnetic and Electric Black Holes in
Arbitrary Dimension},  Phys.Rev. {\bf D80} (2009) 044015.
\bibitem{B1} A. Belhaj, M. Chabab, H. El Moumni, M.B. Sedra, {\em
On non-commutative black holes and their thermodynamics in arbitrary
dimension}, Afr.Rev.Phys. {\bf 8}  (2013) 0017.

\bibitem{x1}A. Belhaj, M. Chabab, H. El Moumni, M. B. Sedra, {\em
Theromodynamics  of AdS Black Holes in Arbitrary Dimensions},
 Chin. Phys. Lett. Vol. {\bf 29}, {No.10}(2012)100401.

\bibitem{x2}  D. Kubiznak, R. B. Mann,  {\em  P-V  criticality of charged AdS black holes},
 JHEP {\bf  1207} (2012)33.

\bibitem{x3}  A. Belhaj,  M. Chabab, L. Medari,  H. El Moumni, M. B. Sedra,  {\em
  The Thermodynamical Behaviors of Kerr�Newman AdS Black Holes}, Chin.Phys.Lett. {\bf 30} (2013)
  090402.
\bibitem{www} S. Ferrara, K. Hayakawa, A. Marrani, {\em Erice Lectures on Black Holes and Attractors},
{\tt arXiv:005.2498 [hep-th]}.  S. Bellucci, S. Ferrara, R. Kallosh,
A. Marrani, {\em  Extremal Black Hole and Flux Vacua Attractors},
Lect. Notes Phys. 755(2008)115-191,  {\tt arXiv:0711.4547 [hep-th]}.
S. Bellucci, S. Ferrara, A. Marrani, A. Yeranyan,  {\em stu Black
Holes Unveiled}, Entropy {\bf 10}(4)(2008)507-555, {\tt
arXiv:0807.3503[hep-th]}.

\bibitem{1}D. Grumiller, W. Kummer, D.V. Vassilevich, {\em Dilaton Gravity in Two
Dimensions},  Phys.Rept.{\bf 369}(2002)327-430, {\tt
hep-th/0204253}.
\bibitem{001} S. Nojiri, S. D. Odintsov, {\em
Quantum dilatonic gravity in (D = 2)-dimensions, (D = 4)-dimensions
and (D = 5)-dimensions}, Int.J.Mod.Phys. {\bf A16} (2001) 1015-1108.

\bibitem{2} J. A. Harvey, A. Strominger, {\em Quantum
aspects of black holes},  {\tt  hep-th/9209055}.

\bibitem{002} D. Y. Chen, Q. Q. Jiang, X. T. Zu,
{\em Fermions tunnelling from the charged dilatonic black holes},
C.Q.G. {\bf 25}(2008) 205022.

\bibitem{3} D. A. Easson,  {\em Hawking radiation of non singular black holes in two dimensions},
JHEP{\bf} 02(2003)037.

\bibitem{4}  J. Ambjorn, J. Jurkiewicz, R. Loll, {\em  Quantum Gravity, or The Art of Building
Spacetime, in Approaches to Quantum Gravity}, ed. D. Oriti,
Cambridge University Press (2006).

\bibitem{006}
D. Grumiller, {\em An action for the exact string black hole},
JHEP0505(2005)028.

\bibitem{0006}
D. Grumiller, R. McNees,  {\em Thermodynamics of Black Holes in Two
(and Higher) Dimensions},  JHEP 0704(2007)074.

\bibitem{5} A. Belhaj, K. Bilal, M. Nach, M.B. Sedra, {\em Solutions and Thermodynamics of Noncommutative
Liouville Black Hole},  Int.J.Geom.Meth.Mod.Phys. {\bf 10} (2013)
1350009 .


\bibitem{6}  C. Germani,
G. P. Procopio, {\em Two-dimensional Quantum Black Holes, Branes in
BTZ and Holography}, Phys.Rev. {\bf D74} (2006) 044012, { \tt
arXiv:hep-th/0605068}.
\bibitem{007} S. Nojiri,  S. D. Odintsov,
{\em Unified cosmic history in modified gravity: from F(R) theory to
Lorentz non-invariant models}, Phys.Rept.{\bf 505}(2011)59-144.

\bibitem{600}
R. R. Caldwell and M. Kamionkowski, Ann. Rev. Nucl. Part. Sci. {\bf
59} (2009)397.

\bibitem{601} T. P. Sotiriou and V.
Faraoni, Rev. Mod. Phys. {\bf 82}  (2010)451-497.

\bibitem{602} S. I. Kruglov,  {\em  Modified arctan-gravity model mimicking cosmological
constant}, {\tt arxiv: 1310.6915.[hep-th]}.

\bibitem{9212}  T. Banks, M. O'Loughlin,   {\em Lagrangians for Two Dimensional Black
Holes}, Phys.Rev. {\bf D48} (1993) 698-706.

\bibitem{9509} S. Naftulin, S.D. Odintsov, { \em On higher-derivative dilatonic gravity in two
dimensions}, Mod.Phys.Lett.A10(1995)2071-2080.

\bibitem{7} E. J. Copeland, J. Ellison, A. Lukas, J. Roberts, {\em
Cosmological Solutions of Low-Energy Heterotic M-Theory},
Phys.Rev.{\bf D73}(2006)086009,  {\tt hep-th/0601173}.

\bibitem{82} R. Maartens, D. Wands, B. A. Bassett, I. P. C. Heard, {\em  Chaotic
inflation on the brane}, {\tt hep-ph/9912464}. \bibitem{83} E.
Papantonopoulos, V. Zamarias, {\em  Chaotic Inflation on the Brane
with Induced Gravity}, JCAP {\bf 10} (2004) 001, {\tt
gr-qc/0403090}.

\bibitem{84} R. Allahverdi, J. Garcia-Bellido, K. Enqvist, A.
Mazumdar, {\em Gauge Invariant MSSM Inflaton}, {\tt hep-ph/0605035}.

\bibitem{85} D. H. Lyth, {\em MSSM Inflation}, {\tt hep-ph/0605283}.

\bibitem{9}  M. Trodden, V.F. Mukhanov, R.H. Brandenberger, {\em A nonsingular two-dimensional black
hole}, Phys. Lett. B 316 (1993) {\bf 483} {\tt hep-th/9305111}.
\bibitem{10} A. Belhaj, P. Diaz, M. Naciri, A.  Segui, {\em  On Brane inflation potentials and black
hole attractors}, Int. J. Mod. Phys. {\bf D17} (2008) 911-920.


\bibitem{land} R. H. Landau, {\em Quantum mechanics: non-relativistic theory}, Pergamon Press, New York
(1977).

\bibitem{c11} R. M. Wald, {\em  General Relativity}, Chicago Univ. Press, Chicago: 1984.

\bibitem{13}  J. B. Hartle,  S.W. Hawking, {\em  Path integral derivation of black hole radiance}, Phys. Rev.
{\bf D13} (1976) 2188.
\bibitem{14}  J. R. B, Mann, {\em Conservation Laws and 2D black holes in dilaton gravity}, {\tt
hep-ph/9206044}.

\bibitem{cc14}  J. Gegenberg,  G. Kunstatter, D. L. Martinez, {\em  Observables for two dimensional black holes},
 {\tt gr-qc/9408015}.

 \bibitem{FRWxxx}  H. Kodama, A. Ishibashi,  O. Seto, {\em Brane World Cosmology-Gauge,  Invariant
  Formalism for Perturbation},   Phys. Rev. {\bf D62} (2000)  064022,   {\tt
  hep-th/0004160}.
  \bibitem{9712} T. Kadoyoshi, S. Nojiri, S. D. Odintsov,
  {\em Four-dimensional cosmology from dilaton coupled quantum matter in two
  dimensions},  Phys.Lett.{\bf B425} (1998) 255-264.



 \bibitem{anomal}  Mariano Cadoni, {\em Trace anomaly and the Hawking effect in generic two-dimensional dilaton gravity theories},   Phys. Rev. {\bf D53} (1996)   4413-4420 ,   {\tt
  gr-qc/9510012}.
\end{thebibliography}
\end{document}